\documentclass[twoside]{ilcws07}
\usepackage[latin1]{inputenc}
\usepackage[dvips]{graphicx,epsfig,color}
\usepackage{wrapfig,rotating}
\usepackage{amssymb,amsmath,array}
\def\lapp{\mathrel{\rlap{\raise.5ex\hbox{$<$}}
                    {\lower.5ex\hbox{$\sim$}}}}
\def\gapp{\mathrel{\rlap{\raise.5ex\hbox{$>$}}
                    {\lower.5ex\hbox{$\sim$}}}}
\begin{document}
\title{Effects of polarisation on study of anomalous $VVH$ interactions at
a Linear Collider}
\author{
Sudhansu~S.~Biswal$^1$, Debajyoti Choudhury$^2$, 
Rohini~M.~Godbole$^1$ and Mamta$^3$
\vspace{.3cm}\\
1- Centre for High Energy Physics, Indian Institute of Science,
Bangalore 560 012, India
\vspace{.1cm}\\
2- Department of Physics and Astrophysics, University of Delhi,
Delhi 110 007, India
\vspace{.1cm}\\
3- Department of Physics and Electronics, S.G.T.B. Khalsa College,
University of Delhi, Delhi 110 007, India
}
\maketitle
\begin{abstract}
We investigate the use of beam polarisation as well as final state
$\tau$ polarisation effects in probing the interaction 
of the Higgs boson with a
pair of heavy vector bosons in the process $e^+ e^- \rightarrow f
\bar f H$, where $f$ is any light fermion. The sensitivity of the 
International Linear Collider (ILC) operating at  $\sqrt s=500$ GeV, to 
such $VVH$($V = W/Z$) couplings is examined in a model independent way. 
The effects of ISR and beamstrahlung are discussed.
\end{abstract}
\section{Introduction}
The particle physics community hopes that the LHC will soon present it
with the signal for the Higgs; but, it is to the ILC that we will have to turn
to for establishing it as {\it the} SM Higgs boson through a precision 
measurement of its properties.
The dominant channel of Higgs production at the ILC, 
{\em viz.} 
$e^+ e^- \rightarrow f \bar f H$ where
$f$ is any light fermion, 
proceeds via the $VVH$ interaction with $V = Z (W)$.
The most general form of the $VVH$ vertex, consistent with 
Lorentz--invariance, can be  written as:
\begin{eqnarray}
\Gamma_{\mu\nu} &=& g_V^{SM}\left[a_V \ g_{\mu\nu}+\frac{b_V}{m_V^2}(k_{1\nu}
k_{2\mu}
 - g_{\mu\nu}  \ k_{1} \cdot k_{2})
+\frac{\tilde{b}_V}{m_V^2} \ \epsilon_{\mu\nu\alpha\beta} \
k_1^{\alpha}  k_2^{\beta}\right]
\label{eq:coup}
\end{eqnarray}
where $k_{i}$'s denote the momenta of the two $V$'s and, at the tree level
in the SM, $a_V = 1$ and $b_V = \tilde b_V = 0$.  
In our analysis we assume $a_V$ to be real and retain terms upto linear order 
in other anomalous parts. In an effective theory, the general structure of  
$VVH$ coupling can be derived from dimension--six operators. 
\section{The Final State and Kinematical cuts}
We choose to work with a Higgs boson of mass 120 GeV and consider its 
detection
in the $b~\bar b$ final state with a branching ratio 0.68. Furthermore, we
assume the detection efficiency of $b$-quark to be 70\%. 
We impose kinematical cuts designed to suppress dominant backgrounds.
Cuts $R_1$ ($R_2$) on the invariant mass of the $f \bar f$ system:
$\left| m_{f\bar f} - M_Z \right| \leq (\geq)  5 \, \Gamma_Z$ , can be used
to enhance (suppress) the effect of the $s$--channel $Z$--exchange diagram. 

Statistical fluctuations in the cross-section or in an asymmetry,
for a given luminosity ${\cal L}$ and fractional systematic error
$\epsilon$ , can be written as:
\begin{eqnarray}
\Delta\sigma = {\sqrt{\sigma_{SM}/{\cal L} + \epsilon^2\sigma_{SM}^2}}
~~{\rm and}~~ (\Delta A)^2 = \frac{1-A^2_{SM}}{\sigma_{SM}{\cal L}} 
+ \frac{\epsilon^2}{2}(1- A^2_{SM})^2.
\end{eqnarray}
We demand that the contribution to the observable coming from the
anomalous parts are less than the statistical fluctuation in these
quantities at a chosen level of significance and study the sensitivity
of a LC to probe them. We choose $\epsilon$ = 0.01, $\cal L$ = 500
fb$^{-1}$ and look for a $3 \sigma$ effect.
Note that in the case of polarisation
asymmetries the total luminosity of $500$ fb$^{-1}$
is divided equally among different polarisation states.
\section{$ZZH$ couplings} 
We construct observables (${\cal O}_{i}$) whose  
behaviour (odd/even) under the discrete transformations $C$, $P$ and $\tilde T$ 
(the pseudo time reversal operator which reverses particle momenta and spins
without interchanging initial and final states) is the same as that 
for a particular operator in the effective Lagrangian. This is
achieved by taking the expectation values of signs of various
combinations of measured quantities such as particle momenta and spins
,$C_i$'s, $i \neq 1$. Some of these combinations are listed in
Table~\ref{tab:corr1}. The  observables are cross-sections and 
various asymmetries with polarised beams and polarised final state $\tau$'s,
which we discuss in the following sections and are also listed in the Table. 
\begin{table}[!h]
\begin{tabular}{|r|l|ccccc|c|c|c|}
\hline
\hline
ID&$C_i$ & {$C$} & {$P$} & {$CP$} & {$\tilde T$} &
{$CP \tilde T$} &
$\begin{array}{c}
\mbox{Observa-}\cr
\mbox{ble}({\cal O}_{i})
\end{array}$
& \mbox{Coupling} \\
\hline
\hline
1 &$ $ & {$+$} & {$+$} & {$+$} & {$+$}
& {$+$} &$\sigma$& {$a_z,\Re(b_z)$}  \\
\hline
2a&{$\vec{P}_{e} \cdot \vec{p}_H$} & {$-$} & {$+$} & {$-$}
& {$+$} & {$-$} &$A_{\rm FB}$& $\Im(\tilde b_z)$ \\
\hline
2b&{$(\vec{P_e}\times \vec{p}_H) \cdot \vec{P}_{f}$}
& {$+$} & {$-$} & {$-$} & {$-$} & {$+$} &$A_{\rm UD}$& $\Re(\tilde b_z)$  \\
\hline
2c&{$[\vec{P_e} \cdot \vec{p}_H] *
[(\vec{P_e}\times \vec{p}_H) \cdot \vec{P}_{f}] $}
& {$-$} & {$-$} & {$+$} & {$-$} & {$-$} &$A_{\rm comb}$& $\Im(b_z)$ \\
\hline
2d&{$[\vec{P_e} \cdot \vec{p}_{f}] *
[(\vec{P_e}\times \vec{p}_H) \cdot \vec{P}_{f}] $}
& {$\otimes$} & {$-$} & {$\otimes$} & {$-$} & {$\otimes$}
&$A{^{\prime}}_{\rm{comb}}$& $\Im(b_z), \Re(\tilde b_z)$  \\
\hline
\hline
\end{tabular}
\caption{\label{tab:corr1} {\em Various possible $C_i$'s, their
discrete transformation properties,the anomalous couplings on
which they provide information along with observables ${\cal O}_i$. 
Symbol $\otimes$ indicates that the
corresponding $C_i$'s do not have any definite transformation property
under $CP$ or $\tilde T$. Here, $\vec{P}_e \equiv \vec{p}_{e^-} -
\vec{p}_{e^+}$ and $\vec{P}_f \equiv \vec{p}_{f} - \vec{p}_{\bar f}$
with $\vec{p}_{e^-}$ ($\vec{p}_{e^+}$) is momentum of initial state
electron (positron) and analogously $\vec{p}_{f}$ ($\vec{p}_{\bar f}$)
is the momentum of final state fermions (anti-fermions).}}
\end{table}
\subsection{Use of Polarised Initial Beams} 
The preferentially axial coupling of the $Z$ boson with the charged leptons 
indicate that initial beam polarisation may affect our observables strongly.
A similar statement also holds for the $W$-contribution to 
$\nu_e \bar \nu_e H$ production. 
In our study, we take $e^-/e^+$ beam polarisations to be 80\% and 60\%
respectively and denote $\cal P$ $\equiv (-,+)$ for 
${\cal P}_{e^-} = -0.8$ and ${\cal P}_{e^+}~=~0.6$. 
The forward-backward (FB) asymmetry in the production of the Higgs boson
with respect to (w.r.t.) the $e^-$ direction
(${\cal O}_{2a}$) is odd under $CP$, even under $\tilde T$ and 
hence can be used to 
probe $\Im(\tilde{b}_{Z})$.  The up-down
(UD) asymmetry (${\cal O}_{2b}$) of the fermion w.r.t the H production plane,
is odd under both $CP$ and $\tilde T$
and hence can constrain $\Re(\tilde{b}_{Z})$. In Table~\ref{tab:beamlimit}
we list the limits of sensitivity on
$\Im(\tilde{b}_{Z})$ and $\Re(\tilde{b}_{Z})$ 
possible with polarised beams for
$E_{cm}$ = 500 GeV. We compare these limits with those
obtained using unpolarised beams~\cite{Biswal:2005fh}. It is
clear from Table~\ref{tab:beamlimit} that  use of longitudinally
polarised beams improves the limit of $\Re (\tilde b_Z)$
and $\Im (\tilde b_Z)$ by a factor of upto 5 or 6.  
This improvement can be traced to the circumvention of 
the vanishingly small vector coupling of electron
to the $Z$ boson.
Our results agree with those of Ref.~\cite{Han:2000mi}
if we remove the kinematical cuts as well as the use of finite
b-tagging efficiency implemented in our analysis.
\begin{table}[!h]
\begin{tabular}{|c|l|c|l|}
\hline\hline
\multicolumn{2}{|c|}{Polarised Beams} &
\multicolumn{2}{|c|}{Unpolarised Beams} \\
\hline
Limits & Observables used & Limits & Observables used\\
\hline
$|\Re(\tilde{b}_{z})| \leq 0.070 $ &
$\begin{array}{l}
{\cal O}_{2b}^{\cal P},~\mbox{R1-cut};\cr
\mu^{-}\mu^{+}H~\mbox{final state}\cr
\end{array}$
&$|\Re(\tilde{b}_{z})| \leq 0.41 $ &
$\begin{array}{l}
A_{UD},~\mbox{R1-cut};\cr
\mu^{-}\mu^{+}H~\mbox{final state}\cr
\end{array}$\\
\hline
$|\Im(\tilde{b}_{z})| \leq 0.0079 $ &
$\begin{array}{l}
{\cal O}_{2a}^{\cal P},~\mbox{R1-cut};~\mu^{-}\mu^{+}H,\cr
q\bar qH ~\mbox{final~states}\cr
\end{array}$
&$|\Im(\tilde{b}_{z})| \leq 0.042 $ &
$\begin{array}{l}
A_{FB},~\mbox{R1-cut};~\mu^{-}\mu^{+}H, \cr
q\bar qH ~\mbox{final~states}\cr
\end{array}$\\
\hline\hline
\end{tabular}
\caption{\label{tab:beamlimit}
{\em Limits on anomalous $ZZH$ couplings from various observables at
$3\sigma$ level with polarised and unpolarised beams, for values
of different parameters as listed in the text.} }
\end{table}
\subsection{Use of Final state $\tau$ Polarization}
Since $\tau$ polarisation can be measured~\cite{Bullock:1992yt,
Hagiwara:1989fn, Godbole:2004mq} using the decay $\pi$ energy
distribution, one can also construct observables using the final state
$\tau$ polarisation  to probe $ZZH$ couplings. To demonstrate this,
we construct, various asymmetries for a sample of (as an example)
left handed $\tau$ in the final state. 
Using the combination $C_{\rm 2c}$ of Table~\ref{tab:corr1} we construct
a mixed polar-azimuthal asymmetry, given by
$A_{\rm{comb}}$ = ($\sigma_{FU} - \sigma_{FD} -
\sigma_{BU} + \sigma_{BD}$)/$\sigma$. 
\begin{wrapfigure}{r}{7.5cm}
\includegraphics[width=7.5cm,clip=true]{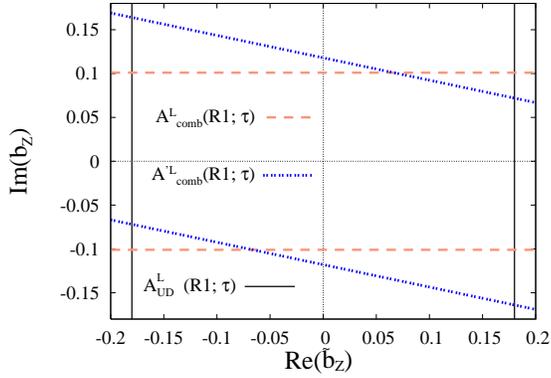}
\caption{\label{fig:T-odd}
{\em Region in $\Re(\tilde b_Z)-\Im(b_Z)$ plane corresponding to the
$3\sigma$ variation of asymmetries with an integrated luminosity of
500 fb$^{-1}$, corresponding to $40\%$ scaling of the asymmetries as 
mentioned in the text.  The horizontal lines are for 3 $\sigma$
variation in $A^L_{\rm{comb}}$, whereas the vertical lines are
for the variation in $A_{\rm{UD}}^L$. The slant lines are corresponding
to variation in $A{^{\prime}}_{\rm{comb}}^L$.}}
\end{wrapfigure}
Here $\sigma$ is total cross section
and $\sigma_{FU}$ is the partial rate with $H$ in the forward(F) hemi-sphere
w.r.t. initial state $e^-$ along with the $\tau^-$ above(U) the $H$ 
production plane etc. It probes $\Im(b_Z)$. 
Similarly we use another combined asymmetry corresponding
to combination $C_{\rm 2d}$, defined as 
$A{^{\prime}}_{\rm{comb}}$ = ($\sigma_{F{^{\prime}}U} - \sigma_{F{^{\prime}}D} -
\sigma_{B{^{\prime}}U} + \sigma_{B{^{\prime}}D}$)/$\sigma$, where
$F^\prime$ ($B^\prime$) corresponds to the production of $\tau^-$ in forward 
(backward) hemi-sphere w.r.t. initial state electron. $U,D$ have the same 
meaning as above.  One may use this asymmetry to constrain both $\Im(b_Z)$ 
and $\Re(\tilde b_Z)$ simultaneously.  The up-down (UD) azimuthal asymmetry
for the  $\tau^-$ can probe $\Re(\tilde b_Z)$. 

The important issue of  efficiency of obtaining a sample enriched with 
$\tau$'s with a particular (say negative) helicity, which we use in the 
analysis, is beyond the scope of discussion here.
Table~\ref{tab:finaltau} lists the limits of sensitivity to different
anomalous couplings, assuming the net effect of having to isolate a negative
helicity $\tau$, to be  just a scaling of asymmetries by $40\%$ and $25 \%$ 
respectively. 
We also compare these with the limits possible without the use of $\tau$ 
polarisation information.  The superscripts $L,1$ in various asymmetries
refer to the helicity of the $\tau$, use of $R1$ cut etc.
Table~\ref{tab:finaltau} shows 
that the use of the  $\tau$ polarisation can improve the sensitivity 
to  $\Im (b_z)$. 
Ref.~\cite{Hagiwara:2000tk} had also pointed out similar improvements on 
using the $\tau$ polarisation in the context of optimal observable analysis. 
Figure~\ref{fig:T-odd} shows the region in $\Re(\tilde b_Z)-\Im(b_Z)$ plane
that can be probed using the above mentioned
asymmetries for $\tau$'s in negative helicity state, 
scaling them by $40\%$ as mentioned earlier.
\begin{table}
\begin{tabular}{||cc||c|c|c||c|c||}
\hline\hline
\multicolumn{5}{||c||}{Using Polarisation of  final state $\tau$} &
\multicolumn{2}{|c||}{Unpolarised Beams} \\
\hline
Coupling & &
\multicolumn{2}{|c|}{Limits}
&
\mbox{Observables~used}
& Limits &
\mbox{Observables~used}
\\
\cline{3-4}
&&
\mbox{40\% eff.}
&
\mbox{25\% eff.}
&
&& \\
\hline
\hline
$|\Im(b_{z})|$
& $\leq$ & 0.10 & 0.13 &
$\begin{array}{l}
A_{1comb}^{L} \cr
\end{array}$
& 0.23 &
$\begin{array}{l}
A_{1comb} \cr
\end{array}$\\
\hline
\hline
$|\Re(\tilde b_{z})|$ & $\leq$ & 0.18 & 0.23 &
$\begin{array}{l}
A_{1UD}^{L} \cr
\end{array}$
& 0.41 &
$\begin{array}{l}
A_{1UD} \cr
\end{array}$\\
\hline
\hline
\end{tabular}
\caption{\label{tab:finaltau} {\em Limits on anomalous $ZZH$ couplings
from various observables at $3 \sigma$ level with/without using the
information of final state $\tau$ polarisation.}}
\end{table}
\section{$WWH$ couplings}
We study the process $e^+ e^- \to \nu \bar \nu H$ with
longitudinally polarised beams to constrain the anomalous $WWH$
couplings. In this case, one can not use the momenta of $\nu$'s to
construct any $\tilde T$-odd observables. We use polarised cross
sections and FB-asymmetry w.r.t. polar angle of the Higgs boson to
probe the anomalous parts of $WWH$ vertex. Keeping only one anomalous
coupling to be nonzero at a time, we obtain individual limits of
sensitivity on these couplings. The values for the same for  $\tilde
T$-odd $WWH$ couplings without/with beam polarisation are listed in
Table \ref{tab:lim-wwh}. 
The simultaneous limits of sensitivity, obtained by letting all the anomalous
couplings to be nonzero, for $\Im(b_W)$ and $\Re(\tilde
b_W)$ with polarised and unpolarised beams are
listed in Table \ref{tab:com-wwh}. It may be noted from the limits
given in Table \ref{tab:lim-wwh} and \ref{tab:com-wwh} that although
use of beam polarisation improves the sensitivity to $\Im(b_W)$ and
$\Re(\tilde b_W)$ by upto a factor 2, there is little reduction in the
contamination coming from the anomalous $ZZH$ couplings.
\begin{table}[!h]
\begin{tabular}{|cc|c|c|c|c|}
\hline
\hline
Coupling & &
$\begin{array}{l}
\mbox{3} \sigma \mbox{ limit with} \cr
\mbox{Polarized} \cr
\mbox{Beams}
\end{array}$
&
$\begin{array}{l}
\mbox{Observable} \cr \mbox{used}
\end{array}$
&
$\begin{array}{l}
\mbox{3} \sigma \mbox{ limit with} \cr
\mbox{Unpolarised} \cr
\mbox{Beams}
\end{array}$
&
$\begin{array}{l}
\mbox{Observable} \cr \mbox{used}
\end{array}$
\\ \hline \hline
$|\Im(b_W)|$ &$\leq$& 0.31 &
$\sigma_{1}^{\cal P}$& 0.62 & $\sigma_{1}$
\\ \hline
$|\Re(\tilde b_W)|$ &$\leq$& 0.76
&$A_{1FB}^{\cal P}$  & 1.6
&$A_{1FB}$ \\
\hline
\hline
\end{tabular}
\caption{\label{tab:lim-wwh} {\em Individual limits on
anomalous $\tilde T$-odd $WWH$ couplings with polarised and unpolarised beams
at $3\sigma$ level at an integrated luminosity of 500 fb$^{-1}$.}}
\end{table}
\begin{table}[!h]
\begin{center}
\begin{tabular}{|cc|c|c|}
\hline
\hline
Coupling & &
$\begin{array}{l}
\mbox{3} \sigma \mbox{ limit with} 
\cr
\mbox{Polarized}~ 
\mbox{Beams}
\end{array}$
&
$\begin{array}{l}
\mbox{3} \sigma \mbox{ limit with} 
\cr
\mbox{Unpolarised}~ 
\mbox{Beams}
\end{array}$
\\ \hline
$|\Im(b_W)|$ &$\leq$& 0.71  & 1.6  \\
$|\Re(\tilde b_W)|$ &$\leq$& 1.7 & 3.2 \\
\hline
\hline
\end{tabular}
\caption{\label{tab:com-wwh} {\em Simultaneous limits on
anomalous $\tilde T$-odd $WWH$ couplings with polarised and unpolarised beams
at $3\sigma$ level at an integrated luminosity of 500 fb$^{-1}$.}}
\end{center}
\end{table}
\section{Sensitivity studies at higher c.m. energies.}
The $s$ and $t$ channel behave differently with increasing energy.
It is therefore interesting to study the energy dependence of the
sensitivity of our observables to the anomalous couplings. We have also
investigated the reach in sensitivity of CLIC to $VVH$ couplings at five 
different c.m. energies, namely at 0.5, 0.8, 1, 1.5 and 3 TeV. We found that
going to higher energy can improve the sensitivity and best possible
sensitivity, for example, for $\Re (\tilde b_Z)$ is obtained
at $\sqrt{s}$ = 1 TeV, with R2-cut. This improvement is upto a factor
of 2 as compared to the analysis made earlier for an ILC operating at
500 GeV c.m. energy \cite{Biswal:2005fh}. 
At higher energies, however, both the initial state radiation (ISR) effect 
as well as the effect of beamstrahlung which causes energy loss of 
the  incoming electron (or positron) due to its interaction with the 
electromagnetic field of the opposite bunch, have to be further taken
into account. Corrections coming from both are sizable and change the rates. 
For example,  at $500$ GeV, the ISR effects change the SM 
contributions by $\lapp 15\%$  whereas the contribution coming from (say) 
$\Re(b_Z)$ changes by about $9\%$; with Beamstrahlung at
(say) 1 TeV these effects are $\sim$ $10\%$ and $20\%$ respectively. 
However,
the effect on the limits for  sensitivity that may be obtained is less drastic 
as these affect both the SM as well as anomalous contribution similarly.
At $1$ TeV, for example, the above mentioned limit changes by 
$15 \%$.
\section{Summary}
Thus we show that use of polarised initial beams can yield higher sensitivity 
to $\Re (\tilde b_Z)$, $\Im (\tilde b_Z)$ and to  both the 
$\tilde T$-odd $WWH$ couplings. 
The limit on $\Im (b_Z)$ can be improved by a factor of
2 to 3 using $\tau$ poalrisation as well, even with pessimistic 
assumptions on the efficiency of the polarisation measurement.
We  also study effect of increasing energy on the sensitivity.
For example, at  $\sqrt{s}=1$TeV one obtains an improvement  by a factor 2, 
which further changes by about $15 \%$  due to  ISR and Beamstrahlung effects.
\begin{footnotesize}

\end{footnotesize}%
\end{document}